\documentclass[aps, prd, showpacs, superscriptaddress, groupedaddress, twocolumn]{revtex4} 
\usepackage{graphicx}    
\usepackage{amssymb}
\usepackage{amsmath}
\usepackage{soul}
\usepackage{subfigure}
\usepackage[pagebackref=false, colorlinks=true]{hyperref}
\hypersetup{
linkcolor=blue,     
citecolor=blue,     
urlcolor=blue}   
\usepackage{soul}
\usepackage{amsmath}
\begin{document}
\title{A bonafide model of structure formation from gravitational collapse}
\author{Koushiki}
\email{koushiki.malda@gmail.com}
\affiliation{International Centre for Space and Cosmology, School of Arts and Sciences, Ahmedabad University, Ahmedabad, GUJ 380009, India}
\author{Dipanjan Dey}
\email{deydipanjan7@gmail.com}
\affiliation{Department of Mathematics and Statistics,
Dalhousie University,
Halifax, Nova Scotia,
Canada B3H 3J5}
\author{Pankaj S. Joshi}
\email{psjcosmos@gmail.com}
\affiliation{International Centre for Space and Cosmology, School of Arts and Sciences, Ahmedabad University, Ahmedabad, GUJ 380009, India}
\date{\today}

\begin{abstract}
This paper explores the cosmological implications of a scalar field with a specific potential, crucial for achieving the final equilibrium state of gravitational collapse. We consider a system with two fluids: minimally coupled matter representing dust-like dark matter and a scalar field acting as dark energy. Our model, akin to the top-hat collapse model, focuses on isolated over-dense regions within a closed FLRW metric, while the background follows a flat FLRW metric. We analyze spacetime configurations where these regions undergo initial expansion followed by contraction, deriving the scalar field potential responsible for their equilibrium state. Our fully relativistic approach offers a comprehensive understanding of stable cosmic over-dense regions, without the need for ad-hoc Newtonian virialization.\\

$\textbf{key words}$: Gravitational collapse, Scalar field collapse, equilibrium condition.

\end{abstract}
\maketitle

\section{Introduction}\label{intro}

Gravitational collapse is an essential phenomenon of nature without which we cannot explain the existence of any celestial object in the sky: be it regular objects like stars and galaxies or exotic objects like singularities. Collapse studies were initiated to study the collapse of pressureless dust clouds \cite{Oppenheimer_1939, Datt_1938}, where the cloud has started from a regular configuration and produced a black hole as its end-state. These results were reproduced numerically as well \cite{Goldswirth:1987, Choptuik:1993, Hod:1997, Vazquez:2022}. It was later shown that for different regular initial configurations, end-state singularities are formed, which are visible to asymptotic observers \cite{PSJ1, PSJ2, PSJ3, Mosani, Mosani2, Mosani3, Mosani4, Mosani5, DeyK1, KK1}. Despite being in contradiction with the cosmic censorship conjecture \cite{Penrose}, these studies have been reproduced by numerical analysis \cite{Shapiro, Shapiro2, Nakamura} as well. 

We talk about objects like singularities and stable regular structures like galaxies and galaxy clusters within the same discussion because there is extensive research to show that all of these objects originate from gravitational collapse and are different from each other due to the differences in their initial configurations.
Though there exists a standard model of the structure formation of our Universe, there are many mysteries that are yet to be solved. Therefore, like the end state of continual gravitational collapse, the structure formation of our universe is also an intriguing, hot research topic.
There are multiple prescriptions for structure formation. One such theory involves perturbing the linear gravity regime to produce density fluctuations or over-dense regions to explain structure formation \cite{Anjan, Anjan2}. Another such prescription is top-hat collapse \cite{Gunn, DDey}. In this model, the structures are generated by over-dense matter bubbles, seeded by pressureless dust. These are assumed to be isolated from the rest of the Universe. While the background is described by a flat FLRW geometry, the over-dense sub-universe is governed by a closed FLRW one. This region, after an initial phase of expansion,  starts collapsing under its gravity. Here the matter component is only dust and hence, there is no pressure to balance the gravitational pull and halting the collapse. In the standard top-hat collapse model, Newtonian virialization is used in an ad-hoc way to halt the collapse to a stable virialized state. At the virialization state, an ensemble of particles satisfies the following condition:
\begin{equation}\label{virial}
    \langle T\rangle_{\tau \to \infty} = -\frac{n}{2}\langle V\rangle_{\tau \to \infty}~, \nonumber
\end{equation}
where, $T$ is the kinetic energy and $V$ is the potential energy. The angled brackets denote their ensemble average and the above expression shows their relationship at asymptotic proper time. Also, the force acting on each particle is central in nature and the associated potential $V(r)\propto r^n$. There have been attempts to generalize this in a relativistic framework by Chandrashekhar and others \cite{Chandra, bonazzola, Cristiane}, but for static and asymptotically flat space-times only. So, this is not useful to equilibrate a collapsing cloud. Hence, the top-hat collapse is a semi-Newtonian approach, at its best. The introduction of non-zero pressure resolves this issue and renders the necessity for the Newtonian virialization obsolete \cite{JMN11, Joshi:2013dva, Bhattacharya:2017chr, Dey:2019fja, Alcubierre:2002, Guzman:2002, Bernal:2006}. Recently, a relativistic definition of equilibrium end state in gravitational collapse is investigated \cite{Dey:2023qdt, DeyK:2023, DEY}. 
In \cite{Dey:2023qdt}, the authors demonstrate how their approach is analogous to top-hat collapse and can explain cosmological structure formation. However, in that paper, the authors do not discuss the potential constituents of matter responsible for the asymptotic equilibrium end state of gravitational collapse, nor do they explore the relativistic equivalent of the Newtonian virialization process. The authors explicitly note that the final equilibrium state may or may not resemble the relativistic counterpart of Newtonian virialization, highlighting the necessity for further investigation in this direction.
In \cite{DeyK:2023}, it is shown that a scalar field having some specific type of potential can lead to the above-mentioned end equilibrium state of gravitational collapse. In that paper, we mathematically derive the required potential of the scalar field, and we do not discuss the cosmological relevance of this type of scalar field. Recently, in \cite{Saha1}, the authors discuss how the dynamics of the over-dense regions of dust-like matter would be influenced by the presence of minimally coupled phantom and quintessence-like scalar fields, where their analysis is relativistic up to the virialization. There is a substantial amount of literature where the influence of dark energy in the structure formation of dark matter at a certain cosmological scale is thoroughly investigated. However, it should be noted that in those works, the authors also employ the Newtonian virialization technique in an ad-hoc manner. 

In the current paper, we address the cosmological significance of the scalar field characterized by a specific potential that leads to the end equilibrium state of gravitational collapse. We consider a system of two fluids consisting of minimally coupled matter and a scalar field with non-zero potential, where the matter is dust-like and represents the dark matter, whereas, the scalar field plays the role of dark energy. Similar to the top-hat collapse model, we consider isolated over-dense regions modeled by closed FLRW metric, whereas, the background is modeled by flat FLRW metric. Both the background and the over-dense regions are seeded by the minimally coupled two-fluid system. Our model examines spacetime configurations where over-dense regions undergo an initial expansion followed by contraction. By employing this model, we derive the scalar field potential responsible for the final equilibrium state of gravitational collapse of these over-dense regions. Our analysis demonstrates that our fully relativistic model offers a comprehensive description of stable cosmic over-dense regions, obviating the need for the ad-hoc introduction of Newtonian virialization as in the conventional top-hat collapse model. 

The structure of this paper is as follows: In Section \ref{top-hat}, we explore the general relativistic equivalent of the top-hat collapse model, utilizing the definition of equilibrium state in gravitational collapse as discussed in our previous paper \cite{DeyK:2023}. Following this, Section \ref{cluster} delves into our model, which involves a two-fluid system comprising the minimally coupled scalar field and dust-like matter that seeds both the internal spacetime of the over-dense region and the background spacetime. In Section \ref{sec3}, we derive the required scalar field potential that leads to the end equilibrium state of the gravitational collapse of the over-dense region. Section \ref{conc} gives a summary of the work presented in this paper. Throughout the paper, we use a system of units in which the velocity of light and the universal gravitational constant (multiplied by $8\pi$), are both set equal to unity.

\section{General relativistic counterpart of top-hat collapse}\label{top-hat}

The over-dense region in the top-hat collapse model comprises pressureless dark matter. Initially, it expands before transitioning to collapse after a turnaround time. Eventually, the over-dense region virializes, resulting in a stable, albeit denser, final state. The appropriate geometry for this collapse is that of an FLRW spacetime, whose metric is:
\begin{equation}\label{genmetric}
    ds^2 = -dt^2 +\frac{a^2(t)}{1-k~r^2}dr^2+R^2(t)~ d\Omega^2.
\end{equation}
This is the only choice of metric because imposing spatial homogeneity and spherical symmetry leaves no other option amongst metrices written in orthonormal coordinates. Needless to say, we are bound to stick to spherical symmetry and spatial homogeneity by observations in large scales. Also, relativistic virialization is only possible if the curvature of the $3$D Riemannian space-like sub-manifold $k = 1$. Hence, our only choice is a closed FLRW geometry. The spatial curvature takes care of the overdensity. This patch with a positive spatial curvature can exist as an over-dense region on a flat FLRW background \cite{Wineberg, Bonga}. This is because these patches are very small compared to the cosmological scales and do not disturb the flatness of the Universe as a whole. 

Now, we define the scale factor as:
\begin{equation}\label{scaling}
    R(t) = r~ a(t),
\end{equation}
where $R(t)$ is the physical radius of the collapsing cloud and $r$ is the co-moving coordinate for the same.
To reproduce the cosmological phenomena of structure formation analytically, one must show that the scale factor increases first and then collapses up to an equilibrium state. Since the relativistic collapse of closed FLRW metric seeded by dust-like matter always leads to a spacetime singularity, one needs to bring other phenomena that can halt the collapse of the over-dense region. As mentioned before, in the Top-hat collapse model, Newtonian virialization is employed during the collapsing phase to reach the stable configuration.

Assuming that the initial value of the scale-factor to be $a_0$, the expansion phase is described by:
\begin{eqnarray}\label{expansion}
  a \in (a_0,a_{max}) \begin{cases}
      
  \dot{a} > 0 \\
  \ddot{a} < 0\,\, .

  \end{cases} 
\end{eqnarray}

At the trurnaround radius $R_{max}$:
\begin{eqnarray}
    \label{turnaound}
    a = a_{max}\begin{cases}
      \dot{a} = 0 \\
  \ddot{a} < 0\,\,   ,
    \end{cases}
\end{eqnarray}

and the collapse phase is described by:

\begin{eqnarray}
    \label{collapsephase}
   a \in (a_{max},a_{e})\begin{cases}
      \dot{a} < 0 \\
  \ddot{a} < 0 ,  
    \end{cases}
\end{eqnarray}
at the end of which, we must reach an equilibrium end-state. In the realm of general relativity, a self-gravitating system can reach an equilibrium state \cite{JMN11, Joshi:2013dva, Bhattacharya:2017chr, Dey:2019fja}. A collapsing cloud can achieve a stable equilibrium if:
\begin{eqnarray}
\lim_{t\to \infty}\dot{R}=\lim_{t\to\infty}\ddot{R}=0\,\, .
\label{eqlb}
\end{eqnarray}
In terms of the scale factor, this condition can be written as:
\begin{equation}\label{asym}
    \lim_{t \to \infty} \Dot{a}= \lim_{t \to \infty} \Ddot{a} =0.
\end{equation}
This implies the collapsing system can only reach an end equilibrium state in asymptotic comoving time. This implies at the equilibrium state:
\begin{eqnarray}
\dot{a}_e(r) =\ddot{a}_e(r) =0\,,
\label{stabc}
\end{eqnarray}
where we use subscript $e$ to denote the equilibrium value of any
quantity as
\begin{equation}\label{assymp}
a_e(r) \equiv \lim_{t \to \infty} a(r,t)\, .    
\end{equation}
Since for a given comoving radius $r$, the scale factor $a(r,t)$ is a smooth monotonically decreasing function of co-moving time $t$, the condition $\dot{a} = \ddot{a} = 0$ may be satisfied in a finite comoving time in some collapsing scenarios. However, that is not a stable equilibrium state. The condition $\dot{a} = \ddot{a} = 0$ at a finite comoving time ($t_e$) does not imply an equilibrium state in general. The equilibrium state at a co-moving time $t_e$ requires \cite{DeyK:2023}:
\begin{eqnarray}
   a(r,t) &>& a_e(r)~ \forall ~ t\in [t_{max},t_e) \\ \nonumber 
   \implies \dot{a} &<& 0 ~ \forall ~ t\in [t_{max},t_e)\,\, ,\nonumber\\
 \text{and}~  a(r,t) &=& a_e(r)~ \forall t\geq t_e \nonumber \\
 \implies \dot{a} &=& \ddot{a} = \dddot{a} =...\nonumber \\&=& a^{(n)} = 0 ~ \forall t\geq t_e \,\, ,
\end{eqnarray}
where $t_{max}$ is the epoch of turn-around where the over-dense cloud starts collapsing under its own gravitation pull. Therefore, if the collapsing system reaches the equilibrium state at a finite comoving time ($t_e$) then the first derivative of the scale factor $a(r, t)$ at that time becomes discontinuous (i.e., $\mathcal{C}^0$ function) which is not possible if we consider $g_{\mu\nu}(t,r)$ is at least $\mathcal{C}^2$. Therefore, for all possible scenarios of gravitational collapse, the equilibrium conditions can only be satisfied at the asymptotic comoving time.

\section{Formation of structures with 
scalar field}\label{cluster}
In the over-dense region, the expansion is explained by a closed FLRW metric. The geometry alone takes care of the same. But in the collapse phase, given by Eq.(\ref{collapsephase}),  matter is clustered to form structures as it interacts with the dark energy, seeded by a scalar field. In this phase, the two minimally interacting fluids: pressureless dust and the dark energy seeded by a scalar field, can be described by a general spherically symmetric metric. This metric is written as:
\begin{equation}\label{gensph}
    ds^2 = -e^{2\nu(r,t)}dt^2 +\frac{R'^2(r,t)}{G(r,t)} dr^2 +R^2(r,t) d\Omega^2, 
\end{equation}
where, $\nu(r,t)$, $R(r,t)$ and $G(r,t)$ are functions of co-moving coordinates $r$ and
$t$. The radial and angular variables are separable and  this metric can be used to describe any 
dynamic system that is spherically symmetric. Due to the presence of the undetermined functions ( $\nu(r,t)$, $R(r,t)$ and $G(r,t)$) of $r$ and $t$, the above metric can describe a large class of collapsing models of matter-fields. Here, a dot and a prime above any function are used to specify a partial derivative
of that function with respect to coordinate time and radius, respectively. With minimal interaction between the matter and the scalar field, the stress-energy tensor of this patch is:
\begin{eqnarray}\label{SE}
    T_{\mu \nu} &=& (T_{\mu \nu})_{DM} +   (T_{\mu \nu})_{\phi} \nonumber \\
    &=& (\rho_{DM}+P_{DM}) u_\mu u_\nu + P_{DM} g_{\mu\nu} \\ \nonumber 
    &+& (\rho_{\phi}+P_{\phi}) u_\mu u_\nu + P_{\phi} g_{\mu\nu}
\end{eqnarray}
where, $u$'s represent the four-velocities of the fluid and unless $u^\mu u_\mu = -1$, then the fluid is anisotropic in the co-moving frame. Hence, for an isotropic fluid in its comoving frame, the energy-density and pressure can be written as:
\begin{eqnarray}\label{se}
    \rho &=& \rho_{DM} + \rho_\phi \nonumber \\
    P &=& P_\phi
\end{eqnarray}
The dynamics of this composite system can be described by the general spherically symmetric metric described in Eq.(\ref{gensph}). We define its scale factor first:
\begin{equation}\label{scale1}
    R(r,t)= r v(r,t),
\end{equation}
where we do not assume spatial homogeniety to begin with. We will see later that spatial homogeniety is necessitated naturally in our desired scheme of evolution. 
For simplicity we consider an isotropic fluid which is observationally reasonable as well. In this case, Einstein equations and conservation of energy equations are:
\begin{eqnarray}
\label{den} 
\rho &=& \frac{F'}{R^2 R'}, \\
\label{press}
P &=& - \frac{\dot{F}}{R^2 \dot{R}}, \\
\label{G}
\dot{G} &=& 2 \frac{\nu'}{R'} \dot{R} G ~,\\
\label{nu}
\nu' &=& -\frac{P'}{\rho +P }~,
\end{eqnarray}
where $F(r,t)$ is the Misner-Sharp mass function, which is defined by:
\begin{equation}\label{Misner}
    F=R \left(1-G + e^{-2\nu} {\dot{R}}^2\right).
\end{equation}
The dynamical space-time described in Eq.(\ref{gensph}) can reach an equilibrium in its collapsing phase \cite{JMN11}. We show later that introducing a scalar field with a specific type of potential halts the collapse. The expansion phase in our model is similar to the expansion phase in standard top-hat collapse, where this phase is dominated by dust. In the collapsing phase, we consider a scalar field minimally coupled to this, whose potential seeds the dark energy effect. \\

As we have discussed in Sec.(\ref{top-hat}), the metric of the over-dense region in the expanding phase is a closed FLRW. So, we match the metrices in Eq.(\ref{genmetric}) and Eq.(\ref{gensph}) at the space-like hypersurface with the maximum scale-factor $(a_{max})$ allowed for the over-dense universe. This is also called the turn-around epoch $t=t_{max}$. It is to be noted that there is nothing sacred about this epoch and it is a choice among many others. We have chosen this because of the relatively lower complexity of zero pressure $\forall ~ t<t_{max}$. According to the Israel-Darmois junction conditions, for the smooth matching at this hypersurface, we must match the first and the second fundamental forms which give us:
\begin{eqnarray}\label{Match}
    \nu (r,t_{max}) &=& 0~, \nonumber \\
    v(r,t_{max}) &=& a_{max} ~, \nonumber\\
    G (r,t_{max}) &=& 1-r^2 ~, \nonumber \\
    v(r,t_{max}) \dot{G} (r,t_{max}) &=& 2 \dot{v} (r,t_{max}) G (r,t_{max}) \nonumber \\ &=&0. 
\end{eqnarray}
Using these junction conditions in Eq.(\ref{gensph}), we get back the closed FLRW metric only.  This is the appropriate choice to describe the collapse of a homogeneous cloud. The flat FLRW metric $(k=0)$ can only describe an unhindered gravitational collapse for a homogeneous cloud whose end-state is a singularity \cite{KK1}. An FLRW spacetime is the appropriate choice as the metric given in Eq.(\ref{gensph}) will become such in a homogeneous case. So, the dynamics described in Sec.(\ref{top-hat}) is completely described by a closed FLRW metric. The co-ordinate interval $0\leq \theta \leq \pi$ gives the expansion phase and $0\leq \theta < \pi$ gives the collapse phase $\forall ~ 0\leq r \leq 1$ \cite{DEY}. Only that the expansion phase is dominated by pressureless dust or dark matter and the effect of dark energy or the scalar field becomes significant post the turn-around epoch $(t>t_{max})$.

 As  $a_{max}$ is not an implicit function of time, $v(r,t_{max})$ is also not a function of time and hence $\dot{G} (r,t_{max}) =0$. From Eq.(\ref{Misner}), the rate of change of the scale factor can be written as:
\begin{equation}\label{scaledot}
    \dot{v}= e^{\nu (r,t)} {\left(\frac{F(r,t)}{r^3 v(r,t)} + \frac{G(r,t)-1}{r^2}\right)}^{\frac12}. 
\end{equation}
In this dynamic model, the expansion goes up to the scale-factor value $a_{max} $ and halts and then starts collapsing. So, at this epoch of turn-around $a=a_{max}$, $\dot{v}$ is zero. Hence, the Misner-Sharp mass function is:
\begin{equation}\label{mass}
    F= v(r,t_{max}) r^3 = a_{max} r^3.
\end{equation}
And with this mass function, the density at this epoch is:
\begin{equation}\label{dense}
    \rho (a_{max}) = \frac{3}{a_{max}^3},
\end{equation}
and this further proves that the fluid is a pressureless dust at this point.
The pressure, can be written with the help of Eq.(\ref{press}), Eq.(\ref{Match}) and Eq.(\ref{mass}) as:
\begin{equation}\label{Press}
    P= \frac{-r\dot{v}(1-G+r^2 \dot{v}^2) -r \dot{G}v +2v\dot{v}r^2\ddot{R}}{r^3\dot{v}v^2},
\end{equation}
and at the turn-around epoch, it becomes:
\begin{equation}\label{Pturn}
    P = \frac{1}{a_{max}^2} + \frac{2\ddot{v} (a_{max})}{a_{max}}.
\end{equation}
In our model, the expansion phase, the cloud is dominated by pressureless dust. Therefore, Eq.(\ref{Pturn}) gives:
\begin{equation}\label{vddot}
    \ddot{v} (a_{max}) = -\frac{1}{2 a_{max}}.
\end{equation}
\\
After the turn-around epoch $t> t_{max} (a<a_{max})$, the cloud starts collapsing and in this phase, the effect of dark energy cannot be ignored. We assume the DM profile to be homogeneous and hence $\rho_m = \frac{\rho_{m0}}{a^3}$ which is of the same order of Eq.(\ref{dense}) and $\rho_{m0}$ is the initial density. We assume $\rho_\phi \propto h(a)$. If the equation of state for the scalar field is $\omega_\phi$, then it is related to the equation of state of the composite system $\omega (a)$ like:
\begin{equation}\label{omega}
    \omega_\phi =\omega \left[ 1+ q \left(\frac{h(a)}{h(a_{max})}\right){\left(\frac{a_{max}}{a}\right)}^3\right]
\end{equation}
where, $q=\frac{\rho_\phi}{\rho_{DM}}\vert_{a_{max}}$ at the turn-around epoch. Therefore, the difference between the expansion and the contraction phases is that expansion phase has $\omega =0$ and the equation of state of the collapse phase is variable.  


\section{Conditions for a stable equilibrium end-state}
\label{sec3}

\subsection{Homogeneous scalar field collapse}

 We consider a two-fluid system of a homogeneous scalar field $\phi (t)$ and regular pressureless dark matter that seeds this space-time in Eq.(\ref{genmetric}) to emulate the collapse phase that reaches an equilibrium. Using Einstein equations, the energy density and pressure can be written as:
\begin{eqnarray}
\label{rho}
    \rho &=&\frac{\Dot{\phi}^2}{2}+V(\phi) + \rho_m =3\left(\frac{\Dot{a}^2}{a^2}+\frac{k}{a^2}\right)\,\, ,\\
    p &=&\frac{\Dot{\phi}^2}{2}-V(\phi)=-2\frac{\Ddot{a}}{a}-\frac{\Dot{a}^2}{a^2}-\frac{k}{a^2}\,\, ,
    \label{p}
\end{eqnarray}
where $V(\phi)$ is the potential of scalar field. It is to be noted that the potential is a free function whose functional form will be calculated at the equilibrium state, where the energy density ($\rho_{e}$) and pressure ($p_{e}$) become,
\begin{eqnarray}
    \rho_{e}&=&\frac{3k}{a_{e}^2}\,\, ,\\
    p_{e} &=& -\frac{k}{a_{e}^2},
\end{eqnarray}
which implies that at the equilibrium state, the equation of state ($\omega_\phi$) of the scalar field should be: $\omega_\phi = -\frac13$. Hence, with $k=1$, $\rho_e>0$ gives us $\rho_e + 3p_e =0$, so our system obeys the strong energy condition. So, this equilibrium is not caused by an exotic matter field. One can also verify that at the equilibrium state:
\begin{eqnarray}
    \dot{\phi}^2_{e} = V_{e}(\phi) = \frac{2k}{a_{e}^2}\,\, ,
\end{eqnarray}
which implies:
\begin{eqnarray}
    (T_{\phi})_{e} = \frac12 \dot{\phi}^2_{e} = \frac12 V_{e}(\phi)\,\, ,
\end{eqnarray}
where $T_\phi$ is the kinetic energy associated to the scalar field. The above relation between the potential energy and the kinetic energy of the scalar field at equilibrium is surprisingly similar to the virialization condition in Newtonian mechanics \cite{DeyK:2023}. It can be easily verified that the equilibrium conditions cannot be achieved by the spatially flat FLRW spacetime, since for that we get trivial solutions: $\dot{\phi}_{e} = 0$ and $V_{e}(\phi) = 0$ which is meaningless.

Along with the two field equations of the scalar field (Eqs.~(\ref{rho}, \ref{p})), one can also write down the following conservation equation or Klein-Gordon for the scalar field:
\begin{eqnarray}
\dot{\rho}_\phi + 3\frac{\dot{a}}{a}\left(\rho_\phi + p_\phi\right) = 0  \implies \ddot{\phi}+ 3\frac{\dot{a}}{a} \dot{\phi} + V_{,\phi} = 0.
\end{eqnarray}
However, the above equation is not independent since it can be derived from Eqs.~(\ref{rho}, \ref{p}). Therefore, we have two independent equations and three unknowns: $\phi (a), V(\phi)$, and $\dot{a}(a)$. Therefore, we have the freedom to choose one free function. Here, we choose the function $\dot{a}(a)$ 
in such a way that the equilibrium conditions are satisfied. We consider following functional form of $\dot{a}(a)$:
\begin{eqnarray}
   \dot{a}(a) = \beta~ \left(f(a) - f(a_e)\right)^\alpha ~~ \forall a \in [a_e , a_{max}]\, , 
   \label{adota}
\end{eqnarray}
where, the free parameters are $\beta < 0$ and $a_{max} > a_e$. Here, we also assume that $f(a) > f(a_e)~ \forall~ a > a_e$ and $f(a)$ should be $\mathcal{C}^\infty~\forall~ a\in [a_e , a_{max}]$ . We can relax the last assumption by considering $\dot{a}(a) = \beta~ \lvert f(a) - f(a_e)\rvert^\alpha$. Since it is a collapsing scenario, we consider $\beta < 0$. The above expression of $\dot{a}(a)$ ensures  $\dot{a}\to 0$ as $a \to a_e$. Using the above expression of $\dot{a}(a)$, we get the following expression of $\ddot{a}(a)$:
\begin{eqnarray}
    \ddot{a}(a) = \alpha\beta^2 f^{\prime}(a)~ \left(f(a) - f(a_e)\right)^{2\alpha - 1} ~~ \forall a \in [a_e , a_{max}]\, ,
    \label{addota}
\end{eqnarray}
which implies $\alpha > \frac12$, since we need $\ddot{a}(a_e) = 0$. As it was discussed previously, the condition $\dot{a}=\ddot{a}=0$ at a finite comoving time does not imply the equilibrium state considering $a(t)$ as a smooth monotonically decreasing function of comoving time. When a collapsing system approaches the equilibrium state asymptotically, not only $\dot{a}$, $\ddot{a}$ tend to zero but all the higher order derivatives of $a$ with respect to the comoving time should show similar behavior. 
Therefore, considering the smooth behavior of $a(t,r)$ for a given value of $r$, we can define the following modified version of the general relativistic equilibrium state:
\begin{eqnarray}
  \lim_{t\to\infty} a^{(n,~0)}(t,r) = 0,~~\forall n\in \mathbb{Z}^+,
  \label{modeqcond}
\end{eqnarray}
where $a^{(n,~0)}(t,r)$ implies $n$th order partial derivative of $a(t,r)$ with respect to comoving time $t$ and zeroth order partial derivative of the same with respect to comoving radius $r$. Here we express the time derivatives of the scale factor as a function of the scale factor (i.e., $a^{(n)}(a)$). So, the above definition (i.e., Eq.~(\ref{modeqcond})) of the general relativistic equilibrium state is very much crucial to differentiate the scenario where the $\dot{a}$ and $\ddot{a}$ become zero in finite comoving time from the scenario where the collapsing system asymptotically reaches an equilibrium state. 
Now, we can write down the following equilibrium condition for the above-mentioned homogeneous collapse:
\begin{eqnarray}
    a^{(n)}(a_e) = 0,~~\forall n\in \mathbb{Z}^+,
\end{eqnarray}
Using the expression of $\dot{a}(a)$ we can write:
\vspace{-2cm}
\begin{widetext}
\begin{eqnarray}\label{seq1}
    \dot{a} &=& g(a) ,\;\;\;\nonumber \\
    \ddot{a} &=& g(a) \partial_a g(a) ,\;\;\;\nonumber \\
    \dddot{a} &=& g^2(a) \partial_a^2 g(a) + g(a) [\partial_a g(a)]^2 ,\;\;\;\nonumber \\
    \ddddot{a} &=& g^3(a) \partial_a^3 g(a) + 4g^2(a) \partial_a g(a) \partial_a^2 g(a) + g(a) [\partial_a g(a)]^3 ,\;\;\;\nonumber \\
    .\nonumber\\
    .\nonumber\\
    .\nonumber\\
    a^{(n+1)} &=&
    \frac{\partial^n}{\partial t^n} g (a) = \sum \frac{n!}{k_1! k_2! k_3! .... k_n!} \partial_a^k [g(a)] \prod_{j=1}^n \left(\frac{\partial_t^j a}{j!}\right)^{k_j},
\end{eqnarray}
\end{widetext}

where, the summation runs through all possible partitions of $n$ and $k= k_1+k_2+....+k_n$ and $n= k_1+2 k_2+....+n k_n$. Here, we have used Fa\'a de Bruno's formula to obtain the higher order derivatives 
 using chain rule recursively. $\partial_a^k g(a)$ remains to be calculated, so, we again use Fa\'a de Bruno's formula and get:
\begin{widetext}
\begin{eqnarray}\label{seq2}
     \partial_a g(a) &=& \beta \alpha [f(a) -f(a_e)]^{(\alpha -1)} g'(a), \nonumber\\
     \partial^2_a g(a) &=& \beta \alpha (\alpha -1) [f(a) -f(a_e)]^{(\alpha -2)} [g'(a)]^2+\beta \alpha [f(a) -f(a_e)]^{(\alpha -1)} g''(a), \nonumber\\
     \partial^3_a g(a) &=& 
      \beta \alpha (\alpha -1)(\alpha -2) [f(a) -f(a_e)]^{(\alpha -3)} [g'(a)]^3+
     2\beta \alpha (\alpha -1) [f(a) -f(a_e)]^{(\alpha -2)} g'(a) g''(a)\nonumber\\
     &+& \beta \alpha [f(a) -f(a_e)]^{(\alpha -1)} g'''(a), \nonumber\\
     .\nonumber\\
     .\nonumber\\
     .\nonumber\\
     \partial_a^k g(a) &=& \beta  \sum  \frac{\alpha!}{(\alpha -l)!} \frac{k!}{l_1!l_2!l_3!....l_n!}[f(a) -f(a_e)]^{(\alpha -l)} \prod_{m=1}^k \left(\frac{\partial_a^m g(a)}{m!}\right)^{l_m}
 \end{eqnarray}
 \end{widetext}

 where, $l_1 + l_2 + ....+l_n = n$ and $l_1 + 2l_2 + ....+nl_n = k$ and the sums also runs over ever possible partition of $k$. Now, putting Eq.(\ref{seq2}) in Eq.(\ref{seq1}), we get the leading order behaviour of $[f(a)-f(a_e)]$ related to $a^{(n+1)}$. If the leading order term, which has the lowest exponent, can be ignored, then all other terms in the summation are also negligible. This can be inferred as $f(a)\to f(a_e)$ as the system approaches equilibrium assymptotically. For the lowest value of the exponent, $l$ has to have the highest value. From Eq.(\ref{seq2}), we can see that $\vert l \vert_{max} =n$ and the corresponding term can be seen from Eq.(\ref{seq1}). In this, the exponent of $[f(a)- f(a_e)]$ is $n\alpha$. So, the lowest possible exponent of $[f(a)- f(a_e)]$ in the $n$th order derivative of the scale factor is $n\alpha + \alpha -n$. Now, we can list the leading order terms for each of the derivatives of the scale factor:
\begin{eqnarray}\label{seq3}
    \dot{a} &=& [f(a)-f(a_e)]^{\alpha},\nonumber\\
    \ddot{a} &=& [f(a)-f(a_e)]^{(2\alpha-1)},\nonumber\\
    \dddot{a} &=& [f(a)-f(a_e)]^{(3\alpha-2)},\nonumber\\
    .\nonumber\\
    .\nonumber\\
    .\nonumber\\
    {a}^{(n+1)} &=& [f(a)-f(a_e)]^{[\alpha(n+1) -n]}.
\end{eqnarray}
So, the cloud cannot reach equilibrium if $\alpha (n+1)-n \leq 0$, as this will lead to blowing up of this leading order term. This will lead to all lower order terms going to non-zero values and destabilize the equilibrium. So, the necessary condition for it to reach equilibrium is:
\begin{equation}\label{equi}
    \lim_{a\to a_e} a^{(n+1)} \to 0 
    \Rightarrow \alpha > \frac{n}{n+1}~.
\end{equation}
We need all higher order derivatives of $a$ to be zero, so this demands:
\begin{equation}\label{equi1}
    \lim_{\substack{a\to a_e \\ n\to \infty}} a^{(n+1)} \to 0 \Rightarrow \alpha \geq  1.
\end{equation}

Therefore, $\alpha \geq 1$ ensures the null values of all the higher order derivatives of $a(t)$ wrt co-ordinate time in the limit $a\to a_e$. Where, $\dot{a}(a)$ has the functional form given in Eq.(\ref{adota}). If $\alpha = \frac{n}{n+1}$ (i.e., $\alpha < 1$) then all  derivatives upto order $a^{(n)}$ become zero, but not $a^{(n+1)}$. This implies that after a finite co-moving time, $a^{(n-1)}$ will become non-zero and lower order derivatives will also become non-zero consequently under its effect. So the collapsing dynamics is not settling down at a stable equilibrium. A positive value of $a^{(n+1)}(a_e)$ implies that the cloud will start expanding after a halt and the negative value of the same indicates that the cloud will again start collapsing after a pause.  If $ \frac{n-1}{n}<\alpha< \frac{n}{n+1}$, all the derivatives ordered from $\dot{a}$ to $a^{(n)}$ would become zero at $a = a_e$. However, all other higher order derivatives of $a(t)$ (i.e., $a^{(n+1)}, a^{(n+2)}, \cdots a^{(\infty)}$) would blow up at $a=a_e$ which is obviously not the stable equilibrium scenario. Therefore, $\alpha \geq 1$ isolates equilibrium scenarios from various other dynamical scenairos where $\dot{a}, \ddot{a}$ may become zero at a certain comoving time but do not imply a stable equilibrium state. Therefore, from now on we are going to consider only  $\alpha \geq 1$.

\subsection{Scalar field potential required for an equilibrium}
Now, with the dynamical behaviour of $a^{(n)} ~ \forall  n\in [1,\infty)$ at our disposal, we can determine the functional forms of the scalar fields and their associated potentials those give rise to such dynamics.
Taking the expressions of $\dot{a}$ and $\ddot{a}$ from Eq.~(\ref{adota}) and Eq.~(\ref{addota}) respectively and putting them in Eq.(\ref{rho}) and Eq.(\ref{p}) we get:
\begin{widetext}
\begin{eqnarray}\label{phi}
    {\phi,_a}^2 &=& \frac{2}{a^2} +\frac{2}{a^2 \beta^2 {[f(a)-f(a_e)]}^{2\alpha}} -\frac{2 f'(a)}{a[f(a)-f(a_e)]} -\frac{\rho_{m0}}{a^3 \beta^2 {[f(a)-f(a_e)]}^{2\alpha}} \;\;\;\; \forall \;\; [a_{max},a_e], \\
    \label{V}
    V(a) &=& \frac{2}{a^2} + \frac{2\beta^2 {[f(a)-f(a_e)]}^{2\alpha}}{a^2}  +\frac{\beta^2 \alpha f'(a) {[f(a)-f(a_e)]}^{2\alpha -1}}{a} -\frac{\rho_{m0}}{a^3} \;\;\;\; \forall \;\; [a_{max},a_e],\\
    \label{omegatot}
    \omega_{tot} (a) &=& \frac{-2a\alpha \beta f'(a) {[f(a)-f(a_e)]}^{2\alpha -1} -\beta^2 {[f(a)-f(a_e)]}^{2\alpha }-1}{3\beta^2 {[f(a)-f(a_e)]}^{2\alpha } +3} \;\;\;\; \forall \;\; [a_{max},a_e].
\end{eqnarray}
\end{widetext}
The above equations show the functional dependence of $\partial_a\phi$, the scalar field potential $V$ and the equation of state of the composite system $\omega_{tot}$ on $a$ for the collapsing dynamics where collapsing system asymptotically reaches the equilibrium state and $a$ asymptotically reaches $a_e$. So, we can safely assume $[f(a)-f(a_e)] \to 0$ and with $\alpha \geq 1$, we can write Eq.(\ref{phi}), Eq.(\ref{V}) and Eq.(\ref{omegatot}) as:
\begin{eqnarray}\label{phiapp}
{\phi,_a}^2 &\approx & \frac{2}{a^2 \beta^2 {[f(a)-f(a_e)]}^{2\alpha}}\nonumber \\
&-&\frac{\rho_{m0}}{a^3 \beta^2 {[f(a)-f(a_e)]}^{2\alpha}}~, \\
\label{Vapp}
    V(a) &\approx & \frac{2}{a^2} -\frac{\rho_{m0}}{a^3}~, \\
    \label{omegatotapp}
    \omega_{tot} &\approx& -\frac13 ~.
\end{eqnarray}
 Comparing this with Eq.(\ref{omega}), one can see that $q \approx 0 $. For this to occur $\rho_\phi \vert_{max}<< \rho_{DM}\vert_{max}$, which supports our initial assumption that the effect of the scalar field starts showing from the turn-around epoch. \\
For $f(a)= a$, the potential of this scalar field is:
\begin{widetext}
\begin{eqnarray}
\text{when}~~ \alpha &=& 1 :\nonumber\\    
V(\phi) &\approx& \frac{2}{a_e^2 \left[ 1+ K \exp{\left(\frac{a_e^\frac32 \beta}{\sqrt{2 a_e -\rho_{m0}}}\phi \right)}\right]^2} - \frac{\rho_{m0}}{a_e^3 \left[ 1+ K \exp{\left(\frac{a_e^\frac32 \beta}{\sqrt{2 a_e -\rho_{m0}}}\phi \right)}\right]^3} ~~\textbf{for}~~ \phi > 0 \,\, ,\nonumber\\
V(\phi) &\approx& \frac{2}{a_e^2 \left[ 1+ K \exp{\left(-\frac{a_e^\frac32 \beta}{\sqrt{2 a_e -\rho_{m0}}}\phi \right)}\right]^2} - \frac{\rho_{m0}}{a_e^2 \left[ 1+ K \exp{\left(-\frac{a_e^\frac32 \beta}{\sqrt{2 a_e -\rho_{m0}}}\phi \right)}\right]^3} ~~\textbf{for}~~ \phi < 0 \label{V1},
\end{eqnarray}
\end{widetext}

where, $K$ is the integration constant. And for $\alpha >0$, $\phi (a)$ is complexer to solve analytically as $\alpha$ grows. And for such cases, writing down $V(\phi)$ is analytically impossible. 

 This expression is similar to multiple Tachyonic potentials \cite{Ashoke, Ashdada}, superimposed on each other. To be noted that these expressions are true very close to $a_e$. The potential functions $V(\phi)$ cannot be calculated analytically in the range $[a_{max}, a_e)$.

\section{Conclusion and discussion}\label{conc}
In this paper, we give the 
relativistic counterpart to cosmological top-hat collapse. We start with an over-dense region, described by a closed FLRW metric, which is initially expanding in the background of a flat FLRW metric. Both the background and the over-dense region are seeded by a two-fluid system consisting of the minimally coupled scalar field and dust-like matter. Here, we do not discuss the expansion dynamics of the background since our concentration is to derive the scalar field potential that can cause an end equilibrium state of the collapse of over-dense patches. This over-dense patch starts collapsing after reaching a maximum value of scale factor $a_{max}$ and then starts collapsing. We show that this phase of the evolution is also described by a closed FLRW metric only. So, we show that the initially expanding over-dense cloud starts collapsing on itself and this is expressed by the geometric description of a closed FLRW metric alone. Then we give a description of the equilibrium end-state
\begin{equation*}
    a^{(n)}\vert_{a\to a_e} =0~~ \forall ~~ n ~\in \mathbb{Z}^+ ,
\end{equation*}
which is just redefining the relativistic equilibrium condition, but the derivative of the scale-factor is not written in terms of the co-ordinate time. Instead, they are written in terms of the scale-factor itself. This end-state does not require any thermodynamic condition to attain the equilibrium. We have discussed in Sec.(\ref{intro}), that the virialization technique adapted in the standard top-hat collapse is not entirely relativistic and hence an ad-hoc approach. In this paper, our approach to attain the stable equilibrium end-state is fully relativistic. \\
The over-dense patch in our scenario is driven by a pressureless and spatially homogeneous matter, also largely known as cold dark matter. At the epoch of turn-around $t_{max}$ or $a_{max}$, when this patch starts collapsing, the effects of a scalar field start showing and as its effect, a non-zero pressure starts building and this only halts the collapse to an equilibrium. We match the pressures of both sides $t<t_{max}$ and $t>t_{max}$ at this epoch and show that this transition is smooth. As discussed before, the choice of this instant to match these two phases is not special and is done for the sake of relative simplicity. We analytically derive a class of potential functions near the equilibrium, which will give rise to such a scenario. Examining the functional form of these functions, we deduce that they resemble Tachyonic potentials. The effective mass of the scalar field is imaginary. However, the Misner-Sharp mass function, a purely geometric property of space-time \cite{Dadhich}, remains real and positive for a closed FLRW space-time. The implications of this on the structures formed are yet unclear, and we postpone further exploration of this aspect for future work.

 \section{Acknowledgement}

DD would like to acknowledge the support of
the Atlantic Association for Research in the Mathematical Sciences (AARMS) for funding the work.


\begin{thebibliography}{99}
\bibitem{Oppenheimer_1939} J. R. Oppenheimer and H. Snyder, \href{https://journals.aps.org/pr/issues/56/5}{Phys. Rev. Journals Archive \textbf{ 56}, 455 (1939).}

\bibitem{Datt_1938} S. Datt, Zs. f. Phys. \textbf{108} 314 (1938).

\bibitem{Goldswirth:1987}
D.S.~Goldswirth and T.~Piran,
\href{https://journals.aps.org/prd/abstract/10.1103/PhysRevD.36.3575}{Phys. Rev. D \textbf{36}, 3575 (1987).}

\bibitem{Choptuik:1993}
M.W.~Choptuik,
\href{https://journals.aps.org/prl/abstract/10.1103/PhysRevLett.70.9}{Phys. Rev. Lett. \textbf{70}, 9 (1993).}

\textcolor{black}{\bibitem{Hod:1997}
S.~Hod and T.~Piran,
\href{https://journals.aps.org/prd/abstract/10.1103/PhysRevD.55.R440}{Phys. Rev. D \textbf{ 55}, R440(R) (1997).}} 

\textcolor{black}{\bibitem{Vazquez:2022}
E.J.~Vázquez and M.~Alcubierre,
\href{https://journals.aps.org/prd/abstract/10.1103/PhysRevD.105.064071}{Phys. Rev. D \textbf{105},  064071 (2022).}} 


\bibitem{PSJ1} P. S. Joshi, I. H. Dwivedi,\href{https://journals.aps.org/prd/abstract/10.1103/PhysRevD.47.5357}{Phys. Rev. D \textbf{47}, 5357 (1993)}

\bibitem{PSJ2} P. S. Joshi, Global Aspects in Gravitation and Cosmology (Clendron Press, Oxford, 1993).

\bibitem{PSJ3} P. S. Joshi, Gravitational Collapse, and Spacetime Singularities (Cambridge University Press, Cambridge, United Kingdom, 2007).
 
 
\bibitem{Mosani} K. Mosani, D. Dey and P. S. Joshi, \href{https://journals.aps.org/prd/abstract/10.1103/PhysRevD.102.044037}{Phys. Rev. D \textbf{102}, 044037 (2020).}


\bibitem{Mosani2}
K.~Mosani, D.~Dey, and P.~S.~Joshi,
\href{https://journals.aps.org/prd/abstract/10.1103/PhysRevD.101.044052}{
Phys. Rev. D \textbf{101}, no.4, 044052 (2020).}
\bibitem{Mosani3}
K.~Mosani, D.~Dey and P.~S.~Joshi,
\href{https://academic.oup.com/mnras/article-abstract/504/4/4743/6253197?redirectedFrom=fulltext&login=false}{
Mon. Not. Roy. Astron. Soc. \textbf{504}, no.4, 4743-4750 (2021).}
\bibitem{Mosani4}
K.~Mosani, D.~Dey, K.~Bhattacharya and P.~S.~Joshi,
\href{https://journals.aps.org/prd/abstract/10.1103/PhysRevD.105.064048}{
Phys. Rev. D \textbf{105}, no.6, 064048 (2022).}
 \bibitem{Mosani5} 
 Karim Mosani, Dipanjan Dey, Pankaj Joshi,  Gauranga Charan Samanta, Harikrishnan Menon, Vaishnavi Patel, \href{http://iopscience.iop.org/article/10.1088/1361-6382/acd97a}{Classical Quantum Gravity, Accepted (2023)}.
\bibitem{DeyK1}
D.~Dey, P.~S.~Joshi, K.~Mosani and V.~Vertogradov,
\href{https://link.springer.com/article/10.1140/epjc/s10052-022-10401-1}{
Eur. Phys. J. C \textbf{82}, no.5, 431 (2022).}

\textcolor{black}{\bibitem{KK1}
K.~Mosani, K., P.~S.~Joshi, J.V.~Trivedi and T.~Bhanja,
\href{https://journals.aps.org/prd/abstract/10.1103/PhysRevD.108.044049}{
Phys. Rev. D \textbf{ 108}, no.4,  044049 (2023).}}

\bibitem{Penrose} R. Penrose, \href{https://journals.aps.org/prl/abstract/10.1103/PhysRevLett.14.57}{Phys. Rev. Lett. \textbf{14}, 57 (1965).}

\bibitem{Shapiro}
S.~L.~Shapiro, S.~A.~Teukolsky
\href{https://www.jstor.org/stable/29774425?seq=1}{American Scientist, \textbf{79}, 4, pp. 330-343 (1991).}

\bibitem{Shapiro2}
I.~T.~Iliev, P.~R.~Shapiro,
\href{https://academic.oup.com/mnras/article/325/2/468/1155673}{MNRS, \textbf{325},2, 468-482 (2001).}

\bibitem{Nakamura}
T.~Nakamura, M.~Shibata, K.~Nakao
\href{https://academic.oup.com/ptp/article/89/4/821/1889068}{Prog.~Theo.~Phys., \textbf{89}, 4, pp. 821–831 (1993).}

\bibitem{Anjan}
A.~A.~Sen, V.~F.~Cardone, S.~Capozziello and A.~Troisi
\href{https://www.aanda.org/articles/aa/abs/2006/46/aa5229-06/aa5229-06.html}{Astro.~and~Astro.Phys., textbf{460}, 1, pp. 29-36, (2006).}

\bibitem{Anjan2}
M.~C.~Bento, O.~Bertolami and A.~A.~Sen
\href{https://journals.aps.org/prd/abstract/10.1103/PhysRevD.70.083519}{Phys. Rev. D \textbf{70}, 083519 (2004).}

\bibitem{Gunn}
J.~E.~Gunn and J.~R.~Gott, III,
\href{https://doi.org/10.1086/151605}
{
Astrophys. J. \textbf{176}, 1-19 (1972).}

\bibitem{DDey}
K.~Bhattacharya, D.~Dey, A.~Mazumdar, and T.~Sarkar,
\href{https://journals.aps.org/prd/abstract/10.1103/PhysRevD.101.043005}
{Phys. Rev. D \textbf{101}, 043005, (2023).}

\bibitem{Chandra}
S.~Chandrasekhar and G.~Contopoulos
\href{https://www.jstor.org/stable/71450}{Proceedings of the National Academy of Sciences of the United States of America, \textbf{49}, 5, pp. 608-613 (1963).}

\bibitem{bonazzola}
E.~Gourgoulhon and S.~Bonazzola
\href{https://iopscience.iop.org/article/10.1088/0264-9381/11/2/015/pdf}{Class. Quantum Grav. \textbf{11} 443 (1994).}

\bibitem{Cristiane}
C.~Vilain
\href{https://www.researchgate.net/publication/234370174_Virial_theorem_in_general_relativity_-_Consequences_for_stability_of_spherical_symmetry}{The Astrophysical Journal,\textbf{227(1)}:307-318 (1978).}

\bibitem{JMN11} 
P. S. Joshi, D. Malafarina, and R. Narayan, 
\href{http://iopscience.iop.org/article/10.1088/0264-9381/28/23/235018/meta}{Class. Quantum Grav. {\bf 28}, 235018 (2011).}

\bibitem{Joshi:2013dva}
P.~S.~Joshi, D.~Malafarina and R.~Narayan,
\href{https://iopscience.iop.org/article/10.1088/0264-9381/31/1/015002}{Class. Quant. Grav. \textbf{31}, 015002 (2014).}


\bibitem{Bhattacharya:2017chr}
K.~Bhattacharya, D.~Dey, A.~Mazumdar and T.~Sarkar,
\href{https://journals.aps.org/prd/abstract/10.1103/PhysRevD.101.043005}{Phys. Rev. D \textbf{101}, no.4, 043005 (2020).}

\bibitem{Dey:2019fja}
D.~Dey, P.~Kocherlakota and P.~S.~Joshi,
\href{https://arxiv.org/abs/1907.12792}{arXiv:1907.12792 [gr-qc].}

\textcolor{black}{\bibitem{Alcubierre:2002}
M.~Alcubierre, F.S.~Guzmán, T.~Matos, D.~Núñez, L.A.U.~López and P.~Wiederhold,
\href{https://iopscience.iop.org/article/10.1088/0264-9381/19/19/314}{
Class. Quantum Grav. \textbf{ 19}, no.19, 5017 (2003).}}

\textcolor{black}{\bibitem{Guzman:2002}
F.S.~Guzmán and L.A.~Ureña-López,
\href{https://journals.aps.org/prd/abstract/10.1103/PhysRevD.68.024023}{
Phys. Rev. D \textbf{68}, 024023 (2003).}}

\textcolor{black}{\bibitem{Bernal:2006}
A.~Bernal and F.S.~Guzmán,
\href{https://journals.aps.org/prd/abstract/10.1103/PhysRevD.74.063504}{
Phys. Rev. D \textbf{74}, 063504 (2006).}}

\bibitem{Dey:2023qdt}
D.~Dey, N.~T.~Layden, A.~A.~Coley and P.~S.~Joshi,
\href{https://journals.aps.org/prd/abstract/10.1103/PhysRevD.108.044046}{Phys. Rev. D \textbf{108}, 044046 (2023).}

\bibitem{DeyK:2023}
D.~Dey, K.~, and P.~S.~Joshi
\href{https://journals.aps.org/prd/abstract/10.1103/PhysRevD.108.104045}{
Phys. Rev. D \textbf{108}, 104045 (2023).}

\bibitem{DEY}
K.~Bhattacharya, D.~Dey, A.~Mazumdar and T.~Sarkar \href{https://journals.aps.org/prd/abstract/10.1103/PhysRevD.101.043005}{Phys. Rev. D \textbf{101}, 043005 (2020).}


\bibitem{Saha1}
P.~Saha, D.~Dey and K.~Bhattacharya,
\href{https://doi.org/10.1103/PhysRevD.108.084025}{Phys. Rev. D \textbf{108}, no.8, 084025 (2023).}

\bibitem{Bonga}
B.~Bonga, B.~Gupt and N.~Yokomizo
\href{https://iopscience.iop.org/article/10.1088/1475-7516/2016/10/031}{JCAP \textbf{10} 031 (2016).}

\bibitem{Wineberg}
S.~L.~Weinberg, Cosmology ( Oxford University Press, Oxford, 2008).

\bibitem{Ashoke} A.~Sen, \href{https://iopscience.iop.org/article/10.1088/1126-6708/2002/04/048}{JHEP \textbf{04} 048 (2002).}

\bibitem{Ashdada} A.~Sen, \href{https://iopscience.iop.org/article/10.1088/1126-6708/2002/07/065}{JHEP \textbf{07} 065 (2002).}

\bibitem{Dadhich} N.~Dadhich, R.~Goswami and C.~Hansraj, \href{https://arxiv.org/abs/2304.10197}{arXiv:2304.10197v1 [gr-qc] (2023).}


\end{thebibliography}
\end{document}